\documentclass[11pt]{article}
\usepackage{graphicx}
\usepackage[usenames,dvipsnames]{color}
\usepackage{url}
\usepackage[colorlinks = true,
            linkcolor = blue,
            urlcolor  = blue,
            citecolor = blue,
            anchorcolor = blue]{hyperref}
\usepackage{multirow} % added

\usepackage[top=0.9in, bottom=0.9in, left = 0.9in, right=0.9in]{geometry}

\graphicspath{{./figures/}}

\def\Title#1{\begin{center} {\LARGE #1 } \end{center}}
\def\Author#1{\begin{center}{ \sc #1} \end{center}}
\def\Address#1{\begin{center}{ \it #1} \end{center}}

\newcommand\snowmass{\begin{center}\rule[-0.2in]{\hsize}{0.01in}\\\rule{\hsize}{0.01in}\\
\vskip 0.1in Submitted to the Proceedings of the US Community Study\\ 
on the Future of Particle Physics (Snowmass 2021)\\ 
\rule{\hsize}{0.01in}\\\rule[+0.2in]{\hsize}{0.01in} \end{center}}

\begin{document}

\Title{The Sanford Underground Research Facility}
\Author{J. Heise}
\Address{630 East Summit Street, Lead, SD 57754 USA}

%\begin{Abstract}
%\end{Abstract}

\snowmass

\def\thefootnote{\fnsymbol{footnote}}
\setcounter{footnote}{0}

%%%%%%%%%%%%%%%%%%%%%%%%%%%%%%%%%%%%%%%%%%%%%%%%%%%%%%%%%%%%%%%%%%%%%%%%%%%%%

\section*{Executive Summary}

The Sanford Underground Research Facility (SURF) has been operating since 2007 as a dedicated scientific laboratory supporting underground research in rare-process physics, as well as offering research opportunities in other disciplines. SURF laboratory facilities include a Surface Campus as well as campuses at the 4850-foot level (1500~m, 4300~m.w.e.\@) that host a range of significant physics experiments, including those studying dark matter, neutrino properties, and nuclear astrophysics topics. SURF is also home to the Long-Baseline Neutrino Facility (LBNF) that will host the international Deep Underground Neutrino Experiment (DUNE). 

\begin{figure}[!htbp]
  \includegraphics*[height=0.99\columnwidth,angle=270]{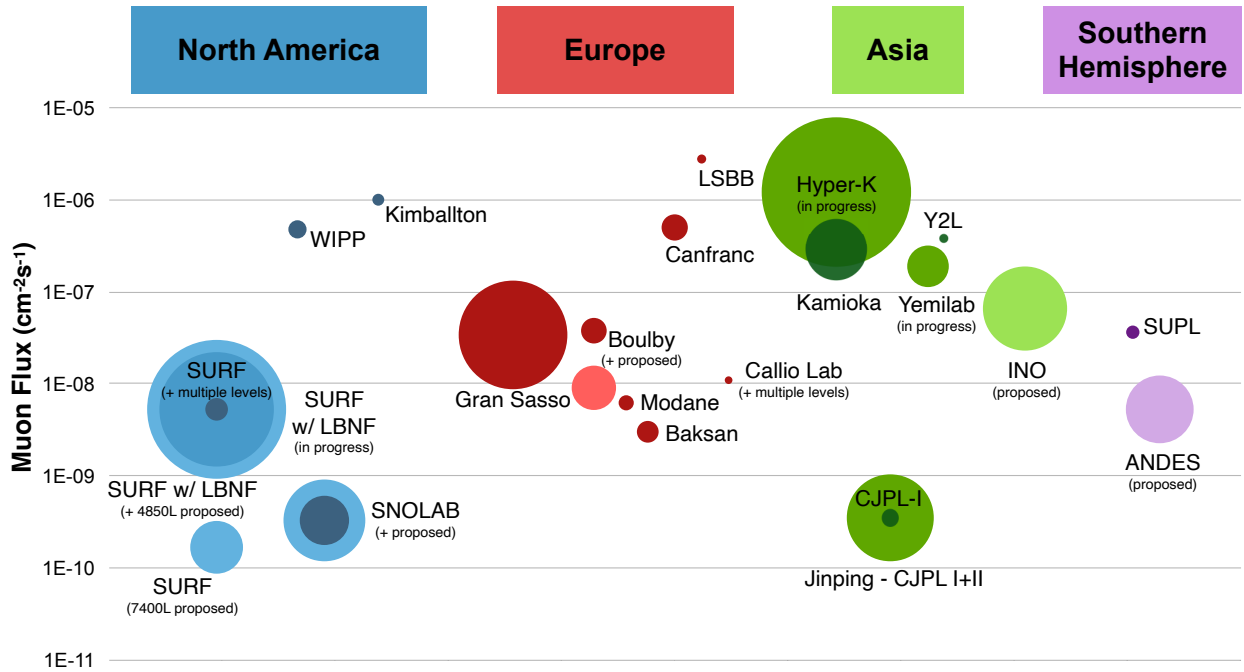}
  \caption{\label{fig:LabVolumes} The size (volume of science space) and effective shielding depth (total muon flux) for the main global underground facilities are represented according to geographic location. The far left-hand side shows the current state and future of SURF. The dark-blue circle represents the current laboratory size, and the surrounding light-blue shaded circle illustrates the expansion once LBNF is realized. The SURF strategic plan aims to provide additional lab spaces on the 4850L (2$\times$ 100-m-long caverns at 1500~m / 4200~m.w.e.\@) as well as a deeper site on the 7400L (2$\times$ 75-m-long caverns at 2300~m / 6500~m.w.e.\@) as indicated by lighter shaded blue circles. Some muon flux values are estimated using a recent parameterization~\cite{JNE:2020bwn}.}
\end{figure}

SURF's capabilities are well-matched to attributes that define a world-class underground facility:

\begin{itemize}
\item Unique environments for multi-disciplinary research: SURF is the deepest underground lab in U.S.\ and one of deepest laboratories in the world, attracting world-leading experiments and scientists from diverse scientific communities. SURF has sufficient depth for next-generation neutrino, rare process and dark matter experiments and is actively exploring expansion opportunities as indicated in \autoref{fig:LabVolumes}.

\item Local radiation shielding: SURF provides a water tank at the Davis Campus and corresponding water purification system. Low-activity facility construction materials were employed in specific areas (e.g., concrete, shotcrete), and in the Davis Cavern additional steel shielding was embedded in the floor below the water tank.

\item Assay capabilities: Low and ultra-low background counting services are available for SURF experiments as well as the international scientific community.

\item Material production/purification: SURF is one of only a few laboratories in the world where underground copper electroforming is currently performed.

\item Environmental control: Cleanrooms with HEPA filtration and dehumidification systems as well as radon-reduction systems (on the surface and underground); some locations have coatings that inhibit radon emanation.

\item Implementation and operations support: SURF has a robust organization with support for experiment planning, installation and operations, with a proven track record of delivering successful science, leveraging and augmented by U.S.\ national laboratory resources as appropriate.

\item Community catalyst: The SURF User Association is serving as a nexus for underground science community planning with recent events such as the SURF Vision Workshop~\cite{SURF_VisionWorkshop}. SURF has also established a Science Program Advisory Committee, and along with the User Association will support the upcoming SURF application to become a DOE Office of Science National User Facility.
\end{itemize}

%%%%%%%%%%% Figure: UG Lab Volumes %%%%%%%%%%
%%%%%%%%%%%%%%%%%%%%%%%%%%%%%%%%%%%%%%%%%%%%%%

As the nation's primary underground laboratory and based on input from the underground science community, SURF advocates for the following recommendations to DOE/NSF prioritizing bodies:
\begin{itemize}
\item Mission need for additional deep laboratory space (including at depths $>$ 6000~m.w.e.\@) in the U.S.\ to support compelling future science.
\item Mission need for a next-generation ($\sim$100-tonne) dark matter and neutrino observatory in the U.S.\
\item Establish a clear and transparent process to optimize the scientific utilization of excavated underground spaces at SURF, including allocating temporary use of a LBNF module as appropriate.
\item Endorse the value of multi-disciplinary underground science at a dedicated laboratory in the U.S.\
\end{itemize}

SURF is a dedicated research facility with significant expansion capability. SURF expansion would enable U.S.\ leadership in many aspects of underground science.

%%%%%%%%%%%%%%%%%%%%%%%%%%%%%%%%%%%%%%%%%%%%%%%%%%%%%%%%%%%%%%%%%%%%%%%%%%%%%%%%%%%

\section{Introduction}

The Sanford Underground Research Facility (SURF) is an international facility dedicated to advancing compelling multidisciplinary underground scientific research, including physics, biology, geology and engineering~\cite{Heise:2021eym,Heise:2017rpu,Heise:2015vza}. The unique underground environment at SURF allows researchers to explore an array of important questions regarding the origin of life and its diversity, mechanisms associated with geologic processes as well as a number of engineering topics such as mining innovations and technology developments. A deep underground laboratory is also where some of the most fundamental topics in physics can be investigated, including the nature of dark matter, the properties of neutrinos and topics related to nuclear astrophysics such as the synthesis of atomic elements within stars. SURF's mission is to advance world-class science and inspire learning across generations.

With strong support from the scientific community as well as federal, state and private (T.\ Denny Sanford) funding, SURF has been operating as a dedicated research facility for 15 years. Since Fall 2011, SURF operation has been funded by the U.S.\ Department of Energy's (DOE) Office of Science, initially via sub-contracts with various national laboratories and through a Cooperative Agreement since Fall 2019.

Since 2006, SURF has provided management and administrative support as well as environment, safety, and health oversight, facility operations and maintenance, science programs and engineering support necessary to host world-leading science experiments at SURF. The SURF organization comprises 203 full/part-time staff in 11 departments and 6 offices. Current and projected SURF staffing levels are indicated in \autoref{tab_personnel}.

%%%%%%%%%%%%%%%%%%%%% Table: Personnel %%%%%%%%%%%%%%%%%%%%%
\begin{table}[htbp]
\caption{\label{tab_personnel} Current and projected SURF staffing levels by category. Full-time equivalent (FTE) personnel as well as the fraction of total staff are indicated.}
\begin{center}
\begin{tabular}{lrrrr}
\hline\hline
{\bf Staffing}    & \multicolumn{2}{c}{\bf FY22}      & \multicolumn{2}{c}{\bf FY27}    \\
{\bf Area}        & \multicolumn{2}{c}{\bf FTE (\%)}  & \multicolumn{2}{c}{\bf FTE (\%)} \\
\hline
Admin/Management  &  21  & (10\%)      &  22 & (10\%)    \\[0.2cm]
Engineering       &  12  &  (6\%)      &  13 &  (6\%)    \\[0.2cm] 
Environment,      &  21  & (10\%)      &  21 &  (9\%)    \\
Safety \& Health  &      &             &     &           \\[0.2cm] 
Outreach          &  20  & (10\%)      &  21 &  (9\%)    \\[0.2cm] 
Scientific        &   6  &  (3\%)      &  11 &  (5\%)    \\[0.2cm]
Technical/        & 123  & (61\%)      & 137 & (61\%)    \\
Operations        &      &             &     &           \\
\hline
TOTAL             & 203  &             & 225 &           \\ 
\hline\hline
\end{tabular}
\end{center}
\end{table}
% Updated Aug 29, 2022: FY27 incl 5 Research Scientists (3 DUNE), + 1 User Facility admin
%%%%%%%%%%%%%%%%%%%%%%%%%%%%%%%%%%%%%%%%%%%%%%%%%%%%%%%

\noindent In particular, SURF provides support for experiment planning, installation and operations that leverages the entire organization. This includes a daily interface with onsite research collaborations to guide activities such as planning and coordination of walkthrough inspections and readiness reviews and related experiment safety evaluations. It also includes evaluating experiments and a process for allocating resources in the context of all SURF activities.

%%%%%%%%%%%%%%%%%%%%%%%%%%%%%%%%%%%%%%%%%%%%%%%%%%%%%%%%%%%%%%%%%%%%%%%%%%%%%%%%%%%

\section{Facilities}

SURF property comprises approximately 1~km$^{2}$ on the surface and more than 31~km$^{2}$ underground. In total, the facility consists of more than 600~km of tunnels extending to over 2450~meters below ground; two main shafts provide redundancy in terms of safe access and some services such as power and network. Considerations for transportation in the shafts are summarized in \autoref{tab:cages}.

%%%%%%%%%%%%%%%%%%%%% Table: Cages %%%%%%%%%%%%%%%%%%%%%
\begin{table}[htbp]
\caption{\label{tab:cages} Summary of key features for the main SURF conveyances and hoists used for transportation of personnel and materials. Cargo loads are transported inside the conveyance and slung loads are transported outside and beneath the conveyance to accommodate larger dimensions but at reduced payload mass. Limits on slung load dimensions correspond to the maximum object height. Additional logistical considerations may be warranted depending on geometry and the underground destination. Note that LBNF/DUNE cryostat beams (longest = 1364~cm) will be transported using the Ross production (skip) hoist and a custom conveyance.}
\begin{center}
\begin{tabular}{lrr}
\hline\hline
{\bf Parameter}      & {\bf Yates} & {\bf Ross} \\
\hline
Personnel, per trip  &   30        &   30 \\
\hline
Width, Cargo (cm)    &  139        &  140 \\
Width, Slung (cm)    &  151        &  142 \\ % Yates was 175
Length, Cargo (cm)   &  377        &  366 \\
Length, Slung (cm)   &  152        &  335 \\ % Yates was 396, Ross was 700
Height, Cargo (cm)   &  258        &  362 \\
Height, Slung (cm)   &  732        &  823 \\ 
\hline
Payload, Cargo (kg)  & 4808        & 6123 \\
Payload, Slung (kg)  & 4536        & 6123 \\
\hline\hline
\end{tabular}
\end{center}
\end{table}
% Access Spec doc, draft Jul 21, 2022
% + Stratman email Mar 9, 2022 
% Updated Aug 2022 with Snowmass CSS presentation values (mainly Ross)
% Max length from Willhite Aug 23, 2022
%%%%%%%%%%%%%%%%%%%%%%%%%%%%%%%%%%%%%%%%%%%%%%%%%%%%%%%

\noindent In general, personnel and materials are transported underground mainly using the Yates cage (LBNF/ DUNE personnel and materials are prioritized at the Ross Shaft). The standard Yates Shaft day-shift schedule affords access 4 days per week up to 10.75 hours per day. Limited periods of 24-hour coverage up to 7 days per week with shifts up to 12.5 hours can be accommodated (shifts beyond 12.5 hours in duration are managed under the SURF fatigue management policy).
% Current Yates Shaft: 6:30 AM - 5:15 PM surface-to-surface -> 10.75 hrs

Research activities at SURF are supported by facilities both on the surface as well as underground. On the surface, the principal facility that directly serves science needs is the Surface Laboratory, which provides approximately 210~m$^{2}$ of lab space (265~m$^{2}$ total). The Surface Laboratory facility includes two cleanrooms (total of more than 90~m$^{2}$), one of which is served by a commercial radon-reduction system capable of a measured reduction of 2200$\times$ at the output and 770$\times$ inside the cleanroom. A new surface maintenance and support facility opened in 2021 that replaces the shipping and receiving warehouse located at the Ross Complex, consolidates maintenance capabilities and resources, provides office space as well as offering some staging space for research groups. The new facility was funded by a \$6.5M state investment and has a total footprint of 2415~m$^{2}$. Existing science storage, staging and assembly space on the surface is summarized in \autoref{tab:storage}. There has been some consideration for future storage, staging and assembly needed to support future science, including experiments hosted at new (non-LBNF/DUNE) underground laboratories. Existing surface buildings could be renovated or new facilities could be constructed on SURF property.

%%%%%%%%%%%%%%%%%% Table: Storage %%%%%%%%%%%%%%%%%%
\begin{table}[htbp]
\caption{\label{tab:storage} Existing surface storage, staging and assembly space at SURF. Some staging areas can be used for non-cleanroom light assembly work. Future storage, staging and assembly needs are also being evaluated.}
\begin{center}
\begin{tabular}{lrl}
\hline\hline
{\bf Use}  & \multicolumn{1}{c}{\bf Footprint}  & {\bf Comment} \\
           & \multicolumn{1}{c}{\bf (m$^{2}$)}  &               \\
\hline
Storage, Cold    & 1385 & Including drill core repository (1015 m$^{2}$) \\ % 14,909 sq.ft, core = 10,924 sq.ft
Storage, Heated  & 220  & Heated space, including some formal HVAC \\ % 2372 sq.ft (Sawmill is heated)
Staging          &  71  & HVAC environment                       \\ % 760 sq.ft
Assembly         & 284  & HVAC environment, some cleanroom space \\ % 3060 sq.ft
\hline\hline
\end{tabular}
\end{center}
\end{table}
% Pam Hamilton: Science Surface Assembly Requirements_21Dec
%%%%%%%%%%%%%%%%%%%%%%%%%%%%%%%%%%%%%%%%%%%%%%%%%%%%%

\begin{figure}[h]
\begin{minipage}{18pc}
\includegraphics*[scale=0.216]{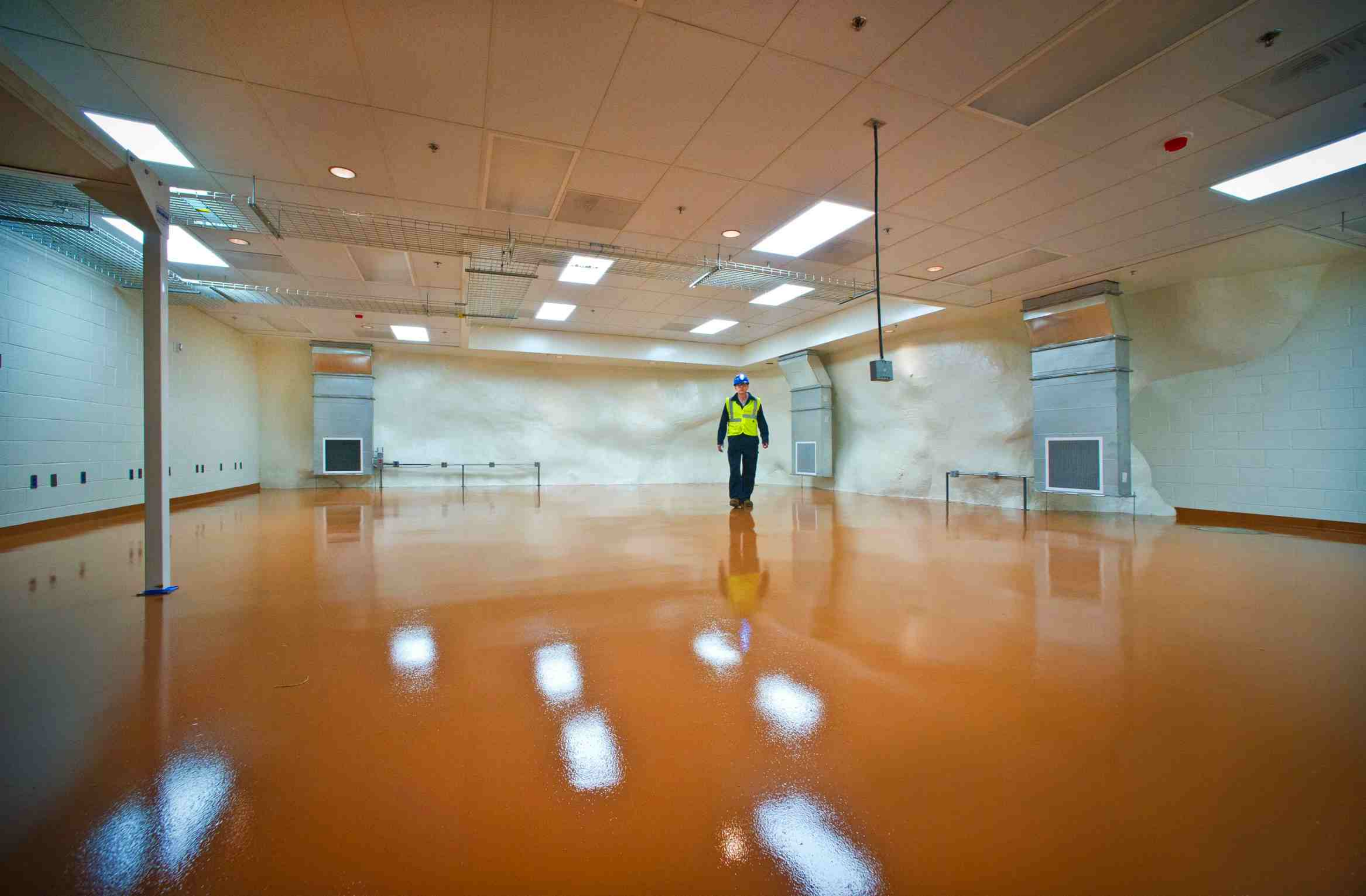}
\end{minipage}\hspace{2.5pc}%
\begin{minipage}{18pc}
\includegraphics*[scale=0.20]{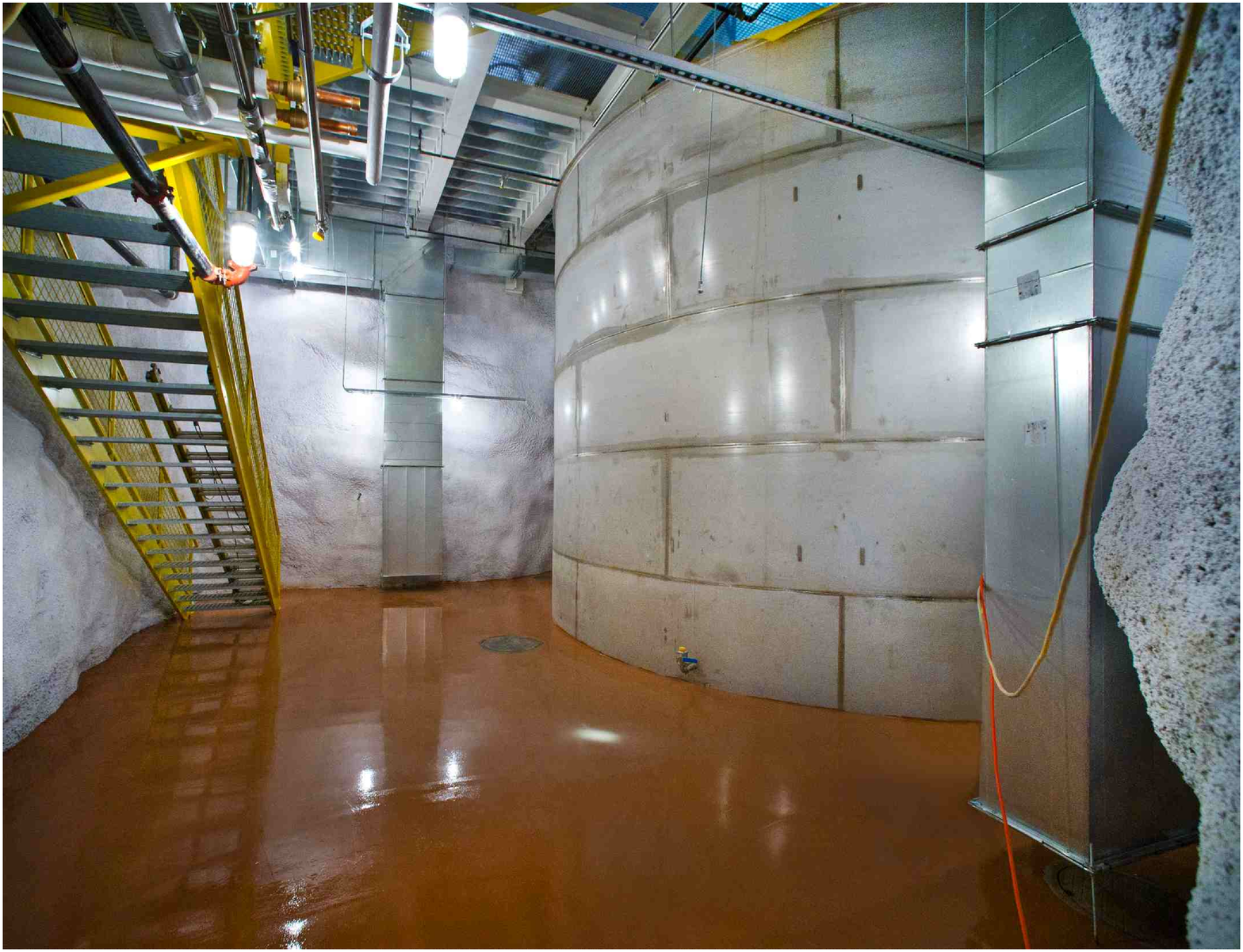}
\end{minipage}
\caption{\label{fig:DavisCampusPics}Pictures of the Davis Campus in May 
2012: (left) Detector Room, (right) Lower Davis Cavern with stainless steel tank.}
\end{figure}

Of the 29 underground elevations currently accessible, areas on seven primary levels have been identified for science activities as summarized in \autoref{tab:science_levels}. Two well-furnished underground research campuses are located on the 4850-foot level of the facility. The Davis Campus (near the Yates Shaft) has a total footprint of 3017~m$^{2}$ and includes a stainless steel tank that can be used for shielding (7.6~m diameter, 6.4~m high); see \autoref{fig:DavisCampusPics}. The Ross Campus (near the Ross Shaft) consists of four areas with a total footprint of 2653~m$^{2}$, with two spaces currently configured as laboratories. See \autoref{tab:footprint} for a summary of footprints for various laboratory spaces. Due to LBNF construction, Ross Campus laboratories were temporarily mothballed in 2021, and activities are expected to resume in $\sim$FY24. Laboratories provide cleanroom spaces as low as class 10--100 with appropriate protocols (see \autoref{tab:cleanrooms}) and are served by redundant utilities, HVAC, access and professional support staff including environment, safety and health, engineering, and scientific support staff. Significant geology and engineering efforts are also underway on the 1700L and 4100L.

%%%%%%%%%%%%%% Table: UG Science Levels %%%%%%%%%%%%%
\begin{table}[htbp]
\caption{\label{tab:science_levels} Summary of key features for the primary SURF underground science levels. For more developed levels, rock overburden details are provided for specific areas such as 4100L alcoves for the EGS Collab -- SIGMA-V project and various 4850L spaces, including the Davis and Ross Campuses as well as the LBNF/DUNE caverns. Characteristics of possible future laboratories on the 4850L (two caverns) and 7400L (one cavern) are also listed in italics. For spaces intended for multiple experiments or detectors, average depth values are presented. Additional overburden density evaluations are available~\cite{Heise:2017rpu}. Note that density-weighted depth values (meters of water equivalent = m.w.e.\@) are determined using a 3-dimensional geological model~\cite{RoggenthenHart-2014}.}
\begin{center}
\begin{tabular}{lrrcl}
\hline\hline
{\bf Science}  & \multicolumn{2}{c}{\bf Vertical Depth} & {\bf Accessible Area} & {\bf Services} \\
{\bf Level}    & {\bf (m)} & {\bf (m.w.e.\@)} & {\bf (Linear distance, m)} & \\
\hline
300L           & 130   &  350    & 1540  & \\
800L           & 280   &  770    & 530   & \\ % 1745' -> 532 m
1700L          &       &         & \multirow{3}{*}{3050} & \\ % 10000' -> 3048 m
\multicolumn{1}{r}{Shafts (avg)} & 550   & 1530   & & \\
\multicolumn{1}{r}{Shops (avg)}  & 330   &  960   & & \\
2000L          & 620   & 1700    & 2970  & Power, network \\ % 9750' -> 2972 m
4100L          &       &         & \multirow{5}{*}{1980} & in limited areas \\ % 6500' -> 1981 m
\multicolumn{1}{r}{Shafts (avg)} & 1280  & 3690   & & \\
\multicolumn{1}{r}{Alcove A}     & 1260  & 3630   & & \\
\multicolumn{1}{r}{Alcove B}     & 1240  & 3560   & & \\
\multicolumn{1}{r}{Powder/Cap}   & 1160  & 3290   & & \\
4550L          & 1430  & 3970    & 1430  & \\ % 4700' -> 1430 m
\hline
4850L          &       &         &       & \\
\multicolumn{1}{r}{Davis Campus (avg)}         & 1470  & 4230 & 3800 & \\ 
\multicolumn{1}{r}{Davis Campus--Lab (MJD)}    & 1480  & 4260 & & \\ 
\multicolumn{1}{r}{Davis Campus--Cavern (LZ)}  & 1470  & 4210 & & \\[0.2cm]
\multicolumn{1}{r}{Ross Campus (avg)}  & 1500  & 4280  & & Significant services \\
\multicolumn{1}{r}{Ross Campus--BHUC}  & 1500  & 4380  & & in labs, power \\
\multicolumn{1}{r}{Ross Campus--Hall (CASPAR)} & 1500  & 4170  & & and network in \\
\multicolumn{1}{r}{Ross Campus--Shop}  & 1500  & 4290  & & other areas \\[0.2cm]
\multicolumn{1}{r}{$\hspace{0.5cm}$LBNF North Cavern (avg)} & 1418  & 3990 & \multirow{2}{*}{(+ 1370 LBNF)} & \\
\multicolumn{1}{r}{LBNF South Cavern (avg)}  & 1390  & 3870 & & \\[0.2cm]
\multicolumn{1}{r}{\it New Labs, Site \#1--North (avg)}  & {\it 1490}   & {\it 4190} & {\it (+ 1140 New \#1)} & \\ % 2410'+1325' (Phase 1) = 3735' -> 1138 m
\multicolumn{1}{r}{\it New Labs, Site \#2--South (avg)}  & {\it 1400}  & {\it 3940} & {\it (+ 1110 New \#2)} & \\ % 2315'+1325' (Phase 1) = 3640' -> 1109 m
\hline
{\it 7400L New Lab} & {\it 2260} & {\it 6460} & {\it 1490} & {\it Significant services} \\
\hline\hline
\end{tabular}
\end{center}
\end{table}
% Updated Aug 12, 2022 with New Labs
% TAUP reference for cone angles
%%%%%%%%%%%%%%%%%%%%%%%%%%%%%%%%%%%%%%%%%%%%%%%%%%%%%%

Specifically, the following utilities are available at SURF:

\begin{itemize}
\item Electrical power: Total capacity = 24,000~kW. Currently available capacity = 20,000~kW (accounting for LBNF/DUNE); FY27 available capacity = 15,000~kW (accounting for LBNF/DUNE).
\item Standby power: 2 diesel generators for the 4850L Davis Campus (300~kW fire \& life safety, 50~kW LZ operations), 1 diesel generator for the 4850L Ross Refuge Chamber (40~kW fire \& life safety).
\item Chilled water: Two redundant refrigerant-to-water chillers (246~kW each) service the Davis Campus with $\sim$70~kW available capacity; an additional facility dehumidifier is also in operation at the Davis Campus. Additional cooling capacity is available through use of industrial water.
\item Purified water: 4850L Davis Campus (reverse osmosis + ultra-filtration) = 37.8~lpm.
\item Compressed air: Ross Shaft/Ross Campus (including Refuge Chamber) = 1100~scfm, Yates Shaft = 528~scfm, 4850L Davis Campus (including Refuge Chamber) = 140~scfm.
\item Network: Redundant network connectivity is available with Internet1 (research) and Internet2 (commodity) services. At the 4850L Davis Campus, network bandwidths are available up to 20~Gbps internally (LAN) and 10~Gbps externally (WAN) using Internet1; Internet2 offers up to 100~Mbps externally and WiFi can reach 1~Gbps. Single-mode fiber throughout the facility supports 100~Gbps, core switches can support 40~Gbps and 100~Gbps optics, and edge routing and firewall infrastructure is limited to 10~Gbps. WiFi is widely available at many underground sites, with access points that support 802.11ac (up to 2.3~Gbps).
\end{itemize}

%%%%%%%%%%%%%% Figure: Davis Campus Pics %%%%%%%%%%%%%%
%%%%%%%%%%%%%%%%%%%%%%%%%%%%%%%%%%%%%%%%%%%%%%%%%%%%%%%%%

%%%%%%%%%%%%%%%%%% Table: Footprint %%%%%%%%%%%%%%%%%%%%%
\begin{table}[htbp]
\caption{\label{tab:footprint} Area and volume footprints for various SURF underground laboratory spaces. Total space includes access tunnels and facility support areas in addition to science laboratory areas. Quantities for proposed spaces are indicated in italics with up to two new laboratory caverns at each site (see \autoref{fig:expansion}). For new 4850L laboratories, only one site is expected to be developed (Site \#1 or \#2).}
\begin{center}
\begin{tabular}{lrrrr}
\hline\hline
{\bf Laboratory}      & \multicolumn{2}{c}{\bf Science}   & \multicolumn{2}{c}{\bf Total}     \\
                      & {\bf Area}      & {\bf Volume}    & {\bf Area}      & {\bf Volume}    \\
                      & {\bf (m$^{2}$)} & {\bf (m$^{3}$)} & {\bf (m$^{2}$)} & {\bf (m$^{3}$)} \\
\hline
4850L Davis Campus             &       1018 &        4633  &      3017  &      11,354   \\
4850L Ross Campus              &        920 &        3144  &      2653  &        8805   \\[0.2cm]
4850L LBNF/DUNE                &       9445 &     191,863  &    17,251  &     242,190   \\[0.2cm]
{\it 4850L New 1$\times$100m Lab (Site \#1--North)} & {\it 2011} & {\it 47,304} & {\it 6135} & {\it 73,589}  \\
{\it 4850L New 2$\times$100m Lab (Site \#1--North)} & {\it 4022} & {\it 94,607} & {\it 8707} & {\it 129,784}  \\
{\it 4850L New 1$\times$100m Lab (Site \#2--South)} & {\it 2011} & {\it 47,304} & {\it 3686} & {\it 73,293}  \\
{\it 4850L New 2$\times$100m Lab (Site \#2--South)} & {\it 4022} & {\it 94,607} & {\it 6391} & {\it 129,274}  \\[0.2cm]
{\it 7400L New 1$\times$75m Lab}      & {\it 3053} & {\it 28,152} & {\it 6733} & {\it 47,910}  \\
{\it 7400L New 2$\times$75m Lab}      & {\it $\sim$4178} & {\it $\sim$42,440} & {\it TBD} & {\it TBD} \\
\hline\hline
\end{tabular}
\end{center}
\end{table}
%%%%%%%%%%%%%%%%%%%%%%%%%%%%%%%%%%%%%%%%%%%%%%%%%%

%%%%%%%%%%%%% Table: Cleanroom Space %%%%%%%%%%%%%
\begin{table}[htbp]
\caption{\label{tab:cleanrooms} Summary of SURF clean spaces. Several underground areas have HVAC systems and established cleanliness protocols that support cleanroom operations. Particle counts are monitored in several locations by experiment collaborations as well as the facility (facility monitoring is denoted by $^{\dagger}$; range of median values shown). Particle count values indicated for Ross Campus locations are representative of nominal cleanroom operations. Median particle counts in common spaces at the Davis Campus are typically 100--200 0.5-$\mu$m-diameter particles per ft$^{3}$.}
\begin{center}
\begin{tabular}{lrrrrrl}
\hline\hline
{\bf Space}  & {\bf Footprint} & {\bf Area} & {\bf Height}   & {\bf Volume} & {\bf Particle} & {\bf Time} \\
             &                 &            &                & & {\bf Count}    & {\bf Frame} \\
             & {\bf (m $\times$ m)} & \bf (m$^{2}$) & {\bf (m)} & {\bf (m$^{3}$)} & {\bf (0.5 $\mu$m} & \\
             &                 &            &                &  & {\bf per ft$^{3}$)} & \\ 
\hline
Surface Laboratory &&&&&& \\
\hline
Standard CR   & 6.6 $\times$ 5.6    &  37 & 2.7              & 300 & 1000     & Available \\
              & (incl entry)        &     & (+ pit: 2.4)     & (+ 32) &          & \\
Reduced-Rn CR & 6.6 $\times$ 8.4    &  55 & 3.3 min/         & 614 & 10--100  & Available \\
              & (incl entry)        &     & 4.4 max          &&          & \\
              &                     &     & (+ pit: 3.2)     & (+ 116) &          & \\[0.2cm]
\hline
\multicolumn{6}{l}{4850L Ross Campus (* Occupancy resuming in FY24)} \\
\hline
Counting CR   & 9.1 $\times$ 6.1    &  56 & 2.4  &  136 & $^{\dagger}$10--150 & Occupied* \\
              & (+ entry)           &     &      &      &              & \\
Bio/Geo CR    & 3.0 $\times$ 6.1    &  18 & 2.4  &   45 & 1000--10,000 & Occupied* \\
              & (+ entry)           &     &      &      &              & \\
Hall          & 30 $\times$ 3 (min) & 236 & 2.8  & 1130 & 100,000      & Occupied* \\[0.2cm]
\hline
\multicolumn{2}{l}{4850L Davis Campus} &&&& \\
\hline
Detector Rm   & 11 $\times$ 9.8--12.8  & 138 & 2.7              & 394 & 100--500   & $\sim$2024 \\
              & (raised section:       &     & (raised          &&            & \\
              & 5.9 $\times$ 5.8)      &     & section: 3.2)    &&            & \\
Machine Shop  & 9.8 $\times$ 5.3       &  52 & 2.7              & 141 & 2000       & Occupied \\
Assay         & 7.3 $\times$ 5.6       &  43 & 2.7              & 118 & $^{\dagger}$10--150 & $\sim$2024 \\
E-forming     & 6.3 $\times$ 8.7       &  53 & 2.7              & 146 & 100--1000  & Occupied \\
Cavern Lower  & 13.7 $\times$ 9.1      & 142 & 6.4              & 948 & 10,000     & $\sim$2027 \\
              & with tank              &     &                  &&            & \\
Compressor Rm & 9.1 $\times$ 4.2       &  33 & 4.3              & 140 & 10,000     & $\sim$2027 \\
Mezzanine     & 3.7 $\times$ 9.1       &  33 & 1.7--1.9         &  68 & 10,000     & $\sim$2027 \\
Cavern Upper  & 17.9 $\times$ 16       & 163 & 4.3              & 801 & $^{\dagger}$1400--4500 & $\sim$2027 \\
\hline\hline
\end{tabular}
\end{center}
\end{table}
% Particle counts: March 2022 Hanhardt analysis
%%%%%%%%%%%%%%%%%%%%%%%%%%%%%%%%%%%%%%%%%%%%%%%%%%%

The main Refuge Chamber at the Ross Campus currently supports a maximum occupancy of 144 people, which will be increased to at least 250 people by mid-2024 (in advance of the LBNF/DUNE construction peak expected in Summer 2026). Additional Refuge Chamber provisions are also available at the Davis Campus to support 39 people.

%%%%%%%%%%%%%%%%%%%%%%%%%%%%%%%%%%%%%%%%%%%%%%%%%%%%%%%%%%%%%%%%%%%%%%%%%%%%%%%%%%%

\section{Facility Characterization}

A geologic model has been constructed to incorporate the complex surface topology as well as the seven main geologic formations (plus Rhyolite) as well as other features that characterize the underground environment~\cite{RoggenthenHart-2014}.

SURF and other groups have collected data characterizing the facility in terms of various radioactive backgrounds. The Davis Campus is hosted in Yates Amphibolite rock, which is relatively low in radioactivity: 0.22~ppm U, 0.33~ppm Th and 0.96\% K. The Poorman rock formation surrounding the Ross Campus is slightly higher in natural radioactivity: 2.58~ppm U, 10.48~ppm Th and 2.12\% K~\cite{Assay-Whitepaper, Assay-Oroville}.  

Long-term underground ambient air radon data have been collected at various locations, and recent averages at both the 4850L Davis and Ross Campuses are approximately 300~Bq/m$^{3}$, with a low baseline of 150~Bq/m$^{3}$. Brief excursions have been observed infrequently at both campuses, typically correlated with maintenance and ventilation changes.  

Other efforts to characterize physics backgrounds in a number of underground areas were carried out by various research groups: muons (800L, 2000L~\cite{Bkgd-Muon}, 4850L Davis Campus: 5.31\thinspace$\pm$\thinspace0.17 $\times$ 10$^{-5}$~muons~m$^{-2}$s$^{-1}$~\cite{Bkgd-Muon_MJD}), thermal neutrons (4850L Davis Campus: 1.7\thinspace$\pm$\thinspace0.1 $\times$ 10$^{-2}$~neutrons~m$^{-2}$s$^{-1}$ \cite{Bkgd-Neutron_Best}) and gamma rays (various~\cite{Bkgd-Gamma}, 4850L Davis Campus: 1.9\thinspace$\pm$\thinspace0.4~gammas~cm$^{-2}$s$^{-1}$~\cite{Bkgd-Gamma_LZ}).

%%%%%%%%%%%%%%%%%%%%%%%%%%%%%%%%%%%%%%%%%%%%%%%%%%%%%%%%%%%%%%%%%%%%%%%%%%%%%%%%%%%

\section{Facility Support Capabilities}

Low-background assays for materials associated with SURF experiments as well as others are managed through the Black Hills State University (BHSU) underground campus (BHUC)~\cite{BHUC-Mount2017}. BHUC currently operates five radioassay (gamma-ray counting) instruments at the Davis Campus, three of which have been fully commissioned and characterized and are actively counting samples; two dual-crystal systems as well as another single-crystal system are expected to come online in 2022. Uranium and thorium sensitivities on the order of 0.1~$\mu$Bq/kg ($\sim$1~ppt) are typical for a two-week counting time, and capabilities are summarized in \autoref{tab:lbc_sensitivities}. The SOLO counter that operated at the Ross Campus has been relocated to surface facilities at BHSU. Local universities have some additional material screening capabilities: ICP-MS (BHSU) and radon-emanation characterization (SD Mines).

%%%%%%%%%%%% Table: LBC Sensitivities %%%%%%%%%%%%%
\begin{table}[htbp]
\caption{\label{tab:lbc_sensitivities} Low-background counter sensitivities for a sample of order $\sim$1~kg and counting for approximately two weeks; see also~\cite{LZ-Assay2020}. ``Davis'' and ``Ross'' indicate the respective 4850L campus of installation. Cooling systems for most detectors were upgraded in 2020 to reduce liquid nitrogen use and associated oxygen deficiency hazards.}
\begin{center}
\begin{tabular}{lcccll}
\hline\hline
{\bf Detector} & {\bf Ge}      & {\bf [U]} & {\bf [Th]} & {\bf Install} & {\bf Status/} \\
{\bf (Group)}  & {\bf Crystal} & {\bf mBq/kg} & {\bf mBq/kg}  & {\bf Date} & {\bf Comment} \\
\hline
Maeve   & 2.2 kg, & 0.1   & 0.1    & Davis: Nov 2020  & Production assays.\\
(LBNL)  & p-type  & (10~ppt) & (25~ppt) & Ross: Nov 2015 & Relocated from \\
        & ($\epsilon$=85\%) & & & Davis: May 2014 & Oroville, old Pb \\
        & & & & & inner shield. \\[0.2cm]

Morgan  & 2.1 kg, & 0.2   & 0.2    & Davis: Nov 2020  & Production assays. \\
(LBNL)  & p-type  & (20~ppt) & (50~ppt) & Ross: Nov 2015 & \\
        & ($\epsilon$=85\%) & & & Davis: May 2015 & \\[0.2cm]

Mordred & 1.3 kg, & 0.7 & 0.7  & Davis: Nov 2020 & Production assays. \\
(USD/   & n-type  & (60~ppt) & (175~ppt) & Ross: Jul 2016 &  Shield access \\
CUBED,  & ($\epsilon$=60\%) &&& Davis: Apr 2013 & upgrade. \\
LBNL)   &&&&& \\[0.2cm]

Dual HPGe    & 2$\times$2.1 kg, & $\sim$0.01 & $\sim$0.01 & Davis: Sep 2020 & Operating.  \\
``Twins''    & p-type & ($\sim$1~ppt) & ($\sim$1~ppt) & Ross: Mar 2018, & Flexible shield \\
(LBNL,BHSU,  & ($\epsilon$=2$\times$120\%) &&& Jul 2017 (initial) & configuration. \\
UCSB)        &&&&& \\[0.2cm]

Ge-IV      & 2.0 kg, & 0.04 & 0.03 & Davis: Fall 2022, & Installation \\
(Alabama,  & p-type  & (3~ppt) & (8~ppt) & Nov 2020 (initial) & underway. \\
Kentucky)  & ($\epsilon$=111\%) &&& Ross: Jul 2018, & Vertical design \\
           &&&& Oct 2017 (initial) & w/ gantry and hoist. \\[0.2cm]

Dual HPGe  & 2$\times$1.1 kg, & $<$0.1 & $<$0.1 & Davis: Feb 2022, & Operating. \\
``RHYM+    & p-type & ($<$10~ppt) & ($<$25~ppt) & Sep 2020 (initial) & BEGe low-E $^{210}$Pb \\
RESN''     & ($\epsilon$=2$\times$65\%) &&&& ($<$2 mBq/kg). \\
(LLNL)     &&&&& \\
\hline\hline
\end{tabular}
\end{center}
\end{table}
%%%%%%%%%%%%%%%%%%%%%%%%%%%%%%%%%%%%%%%%%%%%%%%%%%%%%

Production of electroformed copper is also performed at the facility (average U, Th decay chain $\leq$~0.1~$\mu$Bq/kg). The {\sc Majorana} collaboration has produced electroformed copper since mid-2011, and a total of $\sim$2500~kg of electroformed copper was produced for the {\sc Majorana Demonstrator} (MJD) during a period of approximately 4 years. Three baths currently operate at the Davis Campus, and the installation of a fourth bath is underway, with the possibility to expand to eight baths being evaluated. Electroformed copper is nominally for MJD/LEGEND use, but community requests are possible. 

SURF manages 1.5M liters of xenon purchased through state foundation investments (further purified by the LZ collaboration to remove Kr to $\sim$ppq levels). Liquid nitrogen is available on both the surface and underground to provide boiloff nitrogen purge gas.

SURF holds a Nuclear Regulatory Commission (NRC) broad scope license for radioactive materials, with various gamma-ray and neutron survey instruments and a liquid scintillator counting system (Perkins Elmer Model 4910).

%%%%%%%%%%%%%%%%%%%%%%%%%%%%%%%%%%%%%%%%%%%%%%%%%%%%%%%%%%%%%%%%%%%%%%%%%%%%%%%%%%%

\section{Science Program}

Integral to SURF's institutional mission is the advancement of compelling underground, multidisciplinary research. Science efforts that started in 2007 during re-entry into the facility have grown steadily over the past 15 years. Building on the legacy of the Ray Davis chlorine solar-neutrino experiment~\cite{Cleveland:1998nv} that began in 1965 at the Homestake Mine, 60 groups have conducted underground research programs at SURF at various laboratory elevations ranging from surface to the 5000L. A total of 29 research programs are ongoing, 21 of which are onsite regularly. Not including DUNE, approximately 350 individual researchers are active onsite at SURF from a pool of roughly 600 total experiment collaboration members (since the start of SURF efforts in 2007, 650--700 researchers have been active at SURF). Nine U.S.\ national laboratories are represented among 87 institutions from nine countries.

SURF currently hosts several large experiments, including the LUX-ZEPLIN (LZ) dark matter experiment~\cite{LZ-NIM,LZ-Recent1,LZ-Recent2,LUX-ZEPLIN:2022qhg}, the {\sc Majorana Demonstrator} neutrinoless double-beta decay experiment~\cite{Majorana:2013cem,Majorana:2017csj,Majorana:2016hop,Majorana:2019nbd,Majorana:2022udl}, and the Enhanced Geothermal System (EGS) Collab -- SIGMA-V project~\cite{EGS-Overview2020,EGS-Status2020}; the Compact Accelerator System for Performing Astrophysical Research (CASPAR) nuclear astrophysics experiment~\cite{CASPAR-Robertson2016,CASPAR-Strieder2019,Olivas-Gomez:2022tro,Dombos:2022bph,Shahina:2022xwg} recently completed the first phase of operation. Upcoming is DUNE~\cite{DUNE:2020lwj,DUNE:2022aul}, which will investigate neutrino properties (oscillations, CP violation, mass hierarchy), nucleon decay and supernovae at the 4850L LBNF Campus. In 2021, there were expressions of interest from 17 research groups, including projects showcased in recent community workshops~\cite{CosmicVision2017}.

\begin{figure}[!htbp]
  \includegraphics*[height=0.75\columnwidth,angle=270]{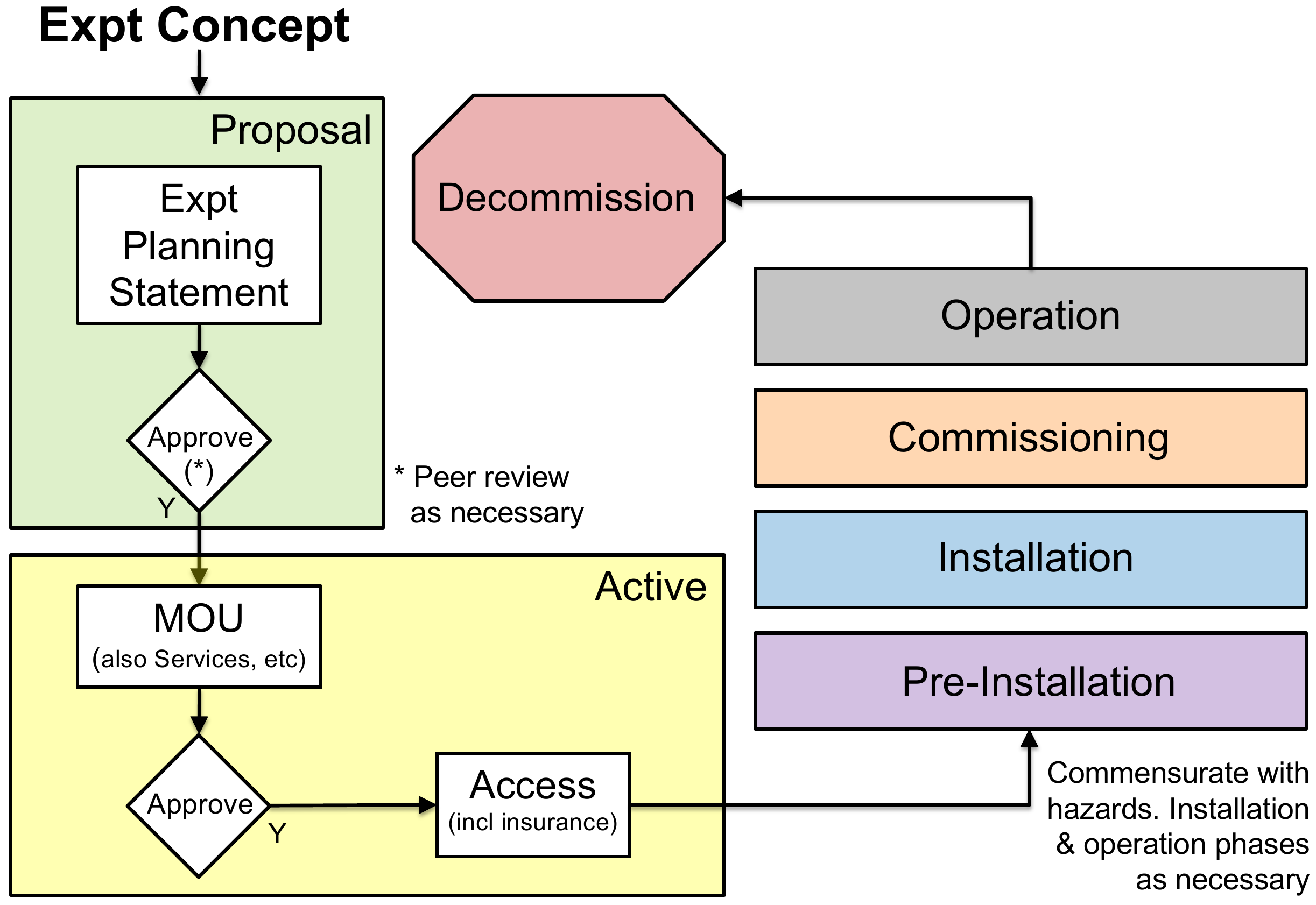}
  \caption{\label{fig:EIP} Flow chart describing elements of the SURF Experiment Implementation Program~\cite{SURF_science}.}
\end{figure}

A formal framework has been in place since 2010 for implementing experiments at SURF in an effective and efficient manner~\cite{SURF_science} as depicted in \autoref{fig:EIP}. In particular, specific documentation helps identify interfaces with the facility, address hazards and establish and define the relationship between an experiment and SURF. Under the DOE Cooperative Agreement, experiments at SURF are offered basic support services, and needs beyond basic services are negotiated on a case-by-base basis.

SURF plans to submit an application in 2022 to become a DOE Office of Science Designated National User Facility.

%%%%%%%%%%%%%%%% Figure: EIP %%%%%%%%%%%%%%%%
% rotated differently between latex and pdflatex, arXiv uses latex
%%%%%%%%%%%%%%%%%%%%%%%%%%%%%%%%%%%%%%%%%%%%%%

All SURF experiment proposals receive internal SURF review based on an Experiment Planning Statement, and groups requesting significant SURF resources or significant changes to the capacities and/or capabilities of the facility may be subject to further external review and evaluation. Peer review is intended to be commensurate with the resources requested. Facility resources requested by some proposals are allocated by the Laboratory Director based on merit review and prioritized using SDSTA-developed criteria such as technical readiness, scientific impact and exploitation of SURF’s unique characteristics. To evaluate experiment proposals, SURF established an external Science Program Advisory Committee in 2021 consisting of domestic and international scientific experts covering the full range of SURF science. 

%%%%%%%%%%%%%%%%%%%%%%%%%%%%%%%%%%%%%%%%%%%%%%%%%%%%%%%%%%%%%%%%%%%%%%%%%%%%%%%%%%%

\section{Community Engagement}

Launched in 2020, the SURF User Association~\cite{SURF_UserAssociation} provides an additional framework for two-way communication on topics important to researchers, fosters a sense of community amongst SURF experiments and researchers, and promotes the scientific case for underground science and its significance to society as well as acting as a channel for advocacy with various representatives. Membership is open to the global underground science community. The User Association is managed by a nine-member Executive Committee elected by association members and appointed according to charter guidelines that ensure diverse representation. The Executive Committee meets at least quarterly and organizes an annual meeting of the general membership. The first annual general meeting was held September 2021. To function effectively, the User Association Executive Committee has a close relationship with SURF management.

Recently, the SURF User Association convened a Long-Term Vision Workshop~\cite{SURF_VisionWorkshop} relevant to both the Snowmass process as well as SURF plans for space and resources over the next several decades. The round table discussion on current and future underground research programs showcased possible uses and interest in underground facilities and identified possible synergistic research opportunities. The discussion spanned physics, geology, biology, engineering as well as additional possible uses for underground space. Some of the key points from the workshop include the following:
\begin{itemize}
\item All Science Disciplines: Significant interest in additional underground space. Additional excavation both scientifically motivated and cost effective (if following LBNF/DUNE) even if precise details on which experiments not worked out yet.
\item LBNF/DUNE: Other experiments may be able to take advantage of LBNF/DUNE neutrino beam at SURF (e.g., Theia~\cite{Theia-Snowmass}). Significant interest in temporary use of LBNF/DUNE cavern space. Need process for engaging with community to identify potential suitable projects.
\item Dark Matter: Generation-3 detector footprint (including shield) $\sim$10--12~m high (a 20~m W $\times$ 24~m H cavern would work). Also quantum sensors for low-mass dark matter (only modest underground space required for some technologies).
\item Neutrinoless Double-Beta Decay: Expect one more generation beyond tonne-scale (20~m W $\times$ 24~m H cavern would work for $\sim$100~tonne detector, gaseous or natural Xe detector may need larger).
\item Nuclear Astrophysics: CASPAR at SURF still relevant even with other underground accelerators.
\item Atom Interferometry: Vertical shaft $\sim$1000-m length, 2.4-m diameter.
\item Quantum Computing: Cosmic rays/radioactivity cause disruption across multiple qubits. Likely do not need deep site. Synergies could help other physics disciplines.
\item Science Support:
\begin{itemize}
\item Long-Term Access: All research disciplines benefit from access afforded by dedicated science lab (DUNE will ensure longevity of SURF).
\item Low-Background Counting: Assay capability important, consider pre-counting radiopure materials and/or maintain underground stockpile of cosmogenically-sensitive materials.
\item Other Physics Support: Copper electroforming performed at SURF (could do more as needed), crystal growth and fabrication could be performed at SURF (not currently), interest in long-term use of SURF-managed xenon (kilotonne quantities likely require new acquisition techniques).
\item Other Capabilities: Onsite machine shop (surface and/or underground), GPS distribution for timing (including underground).
\end{itemize}
\end{itemize}

Following the 2021 SURF Long-Term Vision workshop, there have been discussions about SURF machine shop use by researchers as well as an initial evaluation of shafts to host experiments requiring significant vertical lengths.

%%%%%%%%%%%%%%%%%%%%%%%%%%%%%%%%%%%%%%%%%%%%%%%%%%%%%%%%%%%%%%%%%%%%%%%%%%%%%%%%%%%

\section{Future Plans}

LBNF/DUNE construction is underway at SURF. The excavation phase for two large caverns (each 145~m L $\times$~20~m W~$\times$ 28~m H) and a utility cavern (190~m L $\times$~20~m W~$\times$ 11~m H) started April 2021 and is expected to last approximately three years. Once completed, LBNF/DUNE will comprise a total of 242,190~m$^{3}$ (see \autoref{tab:footprint}).

\begin{figure}[!htbp]
  \centering
  \includegraphics*[height=0.54\columnwidth,angle=0]{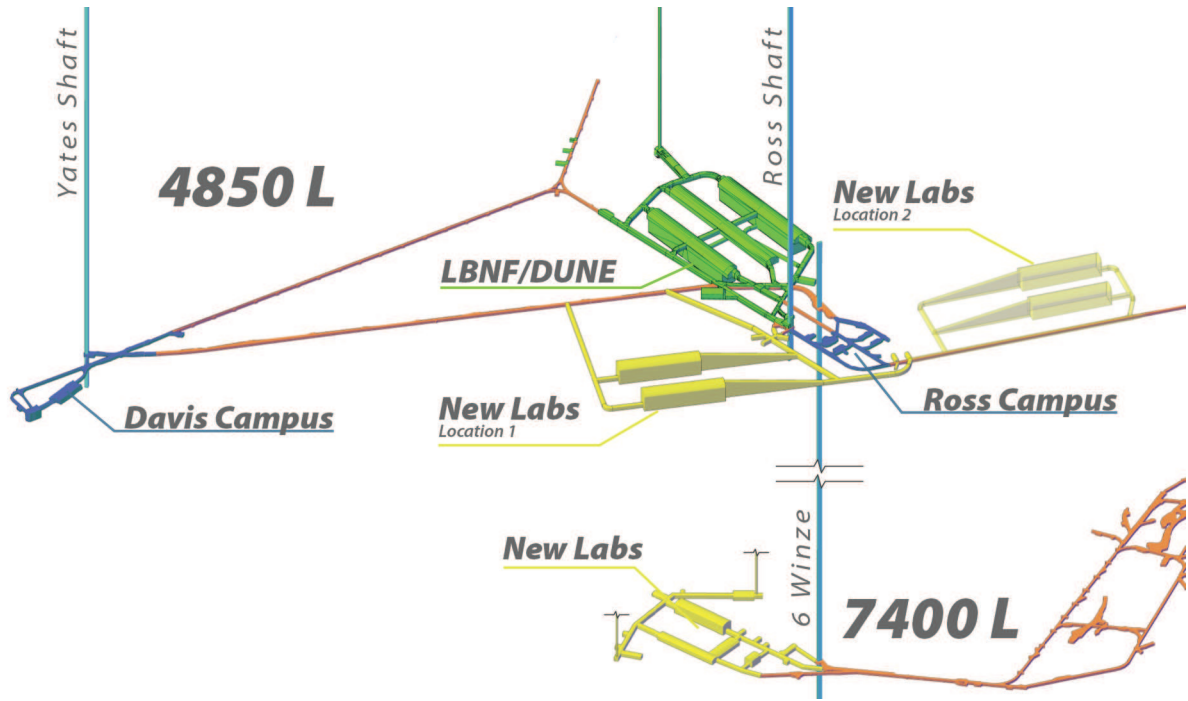}
  \caption{\label{fig:expansion} SURF current and proposed underground laboratory space. Possible locations are shown in yellow for two new caverns (100~m L $\times$~20~m W~$\times$ 24~m H) on the 4850L (1500~m, 4200~m.w.e.\@) as well as new caverns (75~m L $\times$~15~m W~$\times$ 15~m H) on the 7400L (2300~m, 6500~m.w.e.\@); only one 7400L laboratory is shown. Additional caverns or different geometries are also possible at each location.}
\end{figure}

As part of SURF's strategic plan, underground expansion possibilities are being explored. Starting with the 4850L, SURF engaged with an engineering design firm to conduct a feasibility study in 2021 for caverns that could be excavated on a non-interference basis with LBNF/DUNE. Several 4850L locations are viable for laboratories with a cross-section of 20~m wide $\times$ 24~m high and up to 100~m in length, and two of the best location options are depicted in \autoref{fig:expansion} (east-west orientations aligned with the LBNF neutrino beam may also be possible). The average rock overburden for the new 4850L caverns is slightly less than the average overburden of existing 4850L laboratories (4200~m.w.e. versus 4300~m.w.e.\@). There may be advantages to developing laboratories at two separate locations, but each location can accommodate two 100-m caverns, as appropriate. Excavation for one 100-m cavern is estimated to take 2.5~years, including mobilization and de-mobilization, and could begin as early as 2027. The concept for laboratory space on the 7400L (15~m $\times$ 15~m $\times$ 75~m) is based on previous studies~\cite{DUSEL_PDR-Lesko}. Access to the 7400L requires refurbishment of the \#6 Winze as well as development of secondary egress and an additional ventilation pathway.

%%%%%%%%%%%%%%%%%%% Figure: Lab Expansion %%%%%%%%%%%%%%%%%%%
% figure comes in rotated for some reason
%%%%%%%%%%%%%%%%%%%%%%%%%%%%%%%%%%%%%%%%%%%%%%%%%%%%%%%%%%%%%

4850L expansion designs are expected to provide sufficient volume to accommodate future generation dark matter experiments ($\sim$100~tonnes) and neutrinoless double-beta decay projects ($\sim$100~tonnes); see for example~\cite{G3DM-Snowmass}. Other experiments could also take advantage of the LBNF neutrino beam, such as the Theia water-based liquid scintillator project~\cite{Theia-Snowmass}. There is significant opportunity for expansion to meet the needs of a wide range of research disciplines into the future. Several disciplines would benefit from access below the 5000L (1525~m), which is currently the deepest accessible elevation at SURF. These groups include extremophile biology and geothermal projects, and most significantly, physics experiments such as those searching for neutrinoless double-beta decay that are particularly sensitive to cosmogenic backgrounds. The cosmic-ray muon flux on the 7400L (2300~m, 6500~m.w.e.\@) is expected to be 30$\times$ lower than the 4850L and would provide a superior advantage in ensuring cosmogenic backgrounds are negligible.

%%%%%%%%%%%%%%%%%%%%%%%%%%%%%%%%%%%%%%%%%%%%%%%%%%%%%%%%%%%%%%%%%%%%%%%%%%%%%%%%%%%

\section{Summary}

SURF is a deep underground research facility dedicated to scientific uses that has been operating for 15 years and offering world-class service to the underground science community. With a proven track record of enabling experiments to deliver high-impact science, SURF has attracted world-leading experiments and scientists from diverse scientific communities. In addition to the existing science program as well as hosting LBNF/DUNE, SURF is eager to host future experiments.

Research activities are supported at a number of SURF facilities, both on the surface and underground, including cleanrooms and radon-reduction systems, electroformed copper production, as well as robust low-background assay resources.

Expansion possibilities are on the horizon. In addition to the existing two 4850L campuses, SURF is actively exploring options to increase underground laboratory space, and engineering studies have been completed to build large caverns on the 4850L (1500 m, 4200 mwe) and the 7400L (2300~m, 6500~m.w.e.\@). A mixture of federal, state and private funding could allow phased development of underground space aligned with needs for next-generation neutrino and dark matter projects.

%%%%%%%%%%%%%%%%%%%%%%%%%%%%%%%%%%%%%%%%%%%%%%%%%%%%%%%%%%%%%%%%%%%%%%%%%%%%%%%%%%

\section*{Acknowledgments}

This material is based upon work supported by the U.S.\ Department of Energy Office of Science under award number DE-SC0020216.

%%%%%%%%%%%%%%%%%%%%%%%%%%%%%%%%%%%%%%%%%%%%%%%%%%%%%%%%%%%%%%%%%%%%%%%%%%%%%%%%%%

%\bibliographystyle{JHEP}
\bibliographystyle{utphys}
\bibliography{main}

\end{document}